\documentstyle[12pt]{article}

\font \fivesans               = cmss10 at 5pt
\font \sevensans              = cmss10 at 7pt
\font \tensans                = cmss10

\font\boldit=cmbxti10 scaled \magstep1
\newcommand{\stackscript}[1]{\mathop{\smash{#1}\vphantom{#1}}\limits}

\newcommand{\lstop}[1]{\left.}
\newcommand{\rstop}[1]{\right.}

\def\sqr#1#2{{\vcenter{\hrule height.#2pt
    \hbox{\vrule width.#2pt height#1pt \kern#1pt
      \vrule width .#2pt}
    \hrule height.#2pt}}}

\newcounter{exercise}

\def\cor{\mathrel{\mathchoice {\hbox{$\widehat=$}}{\hbox{$\widehat=$}}
{\hbox{$\scriptstyle\hat=$}}
{\hbox{$\scriptscriptstyle\hat=$}}}}

\def\theindex{\@restonecoltrue\if@twocolumn\@restonecolfalse\fi
\columnseprule \z@
\columnsep 35pt\twocolumn[\@makeschapterhead{Index}]
 \@mkboth{INDEX}{INDEX}\thispagestyle{plain}\parindent\z@
 \parskip\z@ plus .3pt\relax\let\item\@idxitem
\addcontentsline{toc}{subsection}{Index}}

\def\bbbr{{\rm I\!R}} %reelle Zahlen

\def\bbbc{{\mathchoice {\setbox0=\hbox{$\displaystyle\rm C$}\hbox{\hbox
to0pt{\kern0.4\wd0\vrule height0.9\ht0\hss}\box0}}
{\setbox0=\hbox{$\textstyle\rm C$}\hbox{\hbox
to0pt{\kern0.4\wd0\vrule height0.9\ht0\hss}\box0}}
{\setbox0=\hbox{$\scriptstyle\rm C$}\hbox{\hbox
to0pt{\kern0.4\wd0\vrule height0.9\ht0\hss}\box0}}
{\setbox0=\hbox{$\scriptscriptstyle\rm C$}\hbox{\hbox
to0pt{\kern0.4\wd0\vrule height0.9\ht0\hss}\box0}}}}
\def\bbbe{{\mathchoice {\setbox0=\hbox{\smalletextfont e}\hbox{\raise
0.1\ht0\hbox to0pt{\kern0.4\wd0\vrule width0.3pt height0.7\ht0\hss}\box0}}
{\setbox0=\hbox{\smalletextfont e}\hbox{\raise
0.1\ht0\hbox to0pt{\kern0.4\wd0\vrule width0.3pt height0.7\ht0\hss}\box0}}
{\setbox0=\hbox{\smallescriptfont e}\hbox{\raise
0.1\ht0\hbox to0pt{\kern0.5\wd0\vrule width0.2pt height0.7\ht0\hss}\box0}}
{\setbox0=\hbox{\smallescriptscriptfont e}\hbox{\raise
0.1\ht0\hbox to0pt{\kern0.4\wd0\vrule width0.2pt height0.7\ht0\hss}\box0}}}}
\def\bbbq{{\mathchoice {\setbox0=\hbox{$\displaystyle\rm Q$}\hbox{\raise
0.15\ht0\hbox to0pt{\kern0.4\wd0\vrule height0.8\ht0\hss}\box0}}
{\setbox0=\hbox{$\textstyle\rm Q$}\hbox{\raise
0.15\ht0\hbox to0pt{\kern0.4\wd0\vrule height0.8\ht0\hss}\box0}}
{\setbox0=\hbox{$\scriptstyle\rm Q$}\hbox{\raise
0.15\ht0\hbox to0pt{\kern0.4\wd0\vrule height0.7\ht0\hss}\box0}}
{\setbox0=\hbox{$\scriptscriptstyle\rm Q$}\hbox{\raise
0.15\ht0\hbox to0pt{\kern0.4\wd0\vrule height0.7\ht0\hss}\box0}}}}
\def\bbbt{{\mathchoice {\setbox0=\hbox{$\displaystyle\rm
T$}\hbox{\hbox to0pt{\kern0.3\wd0\vrule height0.9\ht0\hss}\box0}}
{\setbox0=\hbox{$\textstyle\rm T$}\hbox{\hbox
to0pt{\kern0.3\wd0\vrule height0.9\ht0\hss}\box0}}
{\setbox0=\hbox{$\scriptstyle\rm T$}\hbox{\hbox
to0pt{\kern0.3\wd0\vrule height0.9\ht0\hss}\box0}}
{\setbox0=\hbox{$\scriptscriptstyle\rm T$}\hbox{\hbox
to0pt{\kern0.3\wd0\vrule height0.9\ht0\hss}\box0}}}}
\def\bbbs{{\mathchoice
{\setbox0=\hbox{$\displaystyle     \rm S$}\hbox{\raise0.5\ht0\hbox
to0pt{\kern0.35\wd0\vrule height0.45\ht0\hss}\hbox
to0pt{\kern0.55\wd0\vrule height0.5\ht0\hss}\box0}}
{\setbox0=\hbox{$\textstyle        \rm S$}\hbox{\raise0.5\ht0\hbox
to0pt{\kern0.35\wd0\vrule height0.45\ht0\hss}\hbox
to0pt{\kern0.55\wd0\vrule height0.5\ht0\hss}\box0}}
{\setbox0=\hbox{$\scriptstyle      \rm S$}\hbox{\raise0.5\ht0\hbox
to0pt{\kern0.35\wd0\vrule height0.45\ht0\hss}\raise0.05\ht0\hbox
to0pt{\kern0.5\wd0\vrule height0.45\ht0\hss}\box0}}
{\setbox0=\hbox{$\scriptscriptstyle\rm S$}\hbox{\raise0.5\ht0\hbox
to0pt{\kern0.4\wd0\vrule height0.45\ht0\hss}\raise0.05\ht0\hbox
to0pt{\kern0.55\wd0\vrule height0.45\ht0\hss}\box0}}}}
\def\bbbz{{\mathchoice {\hbox{$\sans\textstyle Z\kern-0.4em Z$}}
{\hbox{$\sans\textstyle Z\kern-0.4em Z$}}
{\hbox{$\sans\scriptstyle Z\kern-0.3em Z$}}
{\hbox{$\sans\scriptscriptstyle Z\kern-0.2em Z$}}}}
\def\qed{\ifmmode\sq\else{\unskip\nobreak\hfil
\penalty50\hskip1em\null\nobreak\hfil\sq
\parfillskip=0pt\finalhyphendemerits=0\endgraf}\fi}
\newfam\sansfam
\textfont\sansfam=\tensans\scriptfont\sansfam=\sevensans
\scriptscriptfont\sansfam=\fivesans
\def\sans{\fam\sansfam\tensans}

\newtheorem{arb}{Working Hypothesis}
\newtheorem{thm}{Theorem}[section]

\newtheorem{lemma}[thm]{Lemma}
\newtheorem{defin}[thm]{Definition}

\newcommand{\hilb}[1]{\mbox{$\cal #1$}}
\newcommand{\algebra}[1]{\mbox{$\cal #1$}}

\newcommand{\region}[1]{\mbox{$\cal #1$}}

\newcommand{\Operator}[1]{\mathchoice
   {\mbox{\boldmath $#1$}}{\mbox{\boldmath $#1$}}
   {\mbox{\footnotesize \boldmath $#1$}}
   {\mbox{\footnotesize \boldmath $#1$}}} %kein \boldmath in \scriptsize

\newcommand{\notion}[1]{\em #1\/\em}
\newcommand{\Notion}[1]{{\boldit #1}}

\def\takeout#1{}
\def\PH{\hilb H\setminus\left\{0\right\}}
\def\SQL{P(\hilb H)}
\def\nlSQL{P_{\algebra A}(\hilb H)}

\def\trace{{\rm tr}}

\def\SSP{S(\hilb H)}
\def\SS{\hilb T(\hilb H)}

\def\sqr#1#2{{\vcenter{\hrule height.#2pt
    \hbox{\vrule width.#2pt height#1pt \kern#1pt
      \vrule width .#2pt}
    \hrule height.#2pt}}}

\def\modulus#1{\left| #1 \right|}
\def\norm#1{\left\| #1 \right\|}

\def\Nv{{\vec \nabla}}

\def\OA{{\Operator A}}

\def\Op#1{{\Operator #1}}
\def\OU{{\Operator U}}

\def\pv{{\vec p}}

\def\xv{{\vec x}}

\normalbaselines
\begin{document}
\input prepictex
\input  pictex
\input postpictex

\hfill ASI-TPA/4/95
\begin{center}
\vspace*{1.0cm}

{\LARGE{\bf Nonlinear Schr\"odinger Dynamics\\[2mm]
and Nonlinear Observables}}

\vskip 1.5cm

{\large {\bf W. L\"ucke}}

\vskip 0.5 cm

Arnold Sommerfeld Institute for
Mathematical Physics\\
Technische Universit\"at Clausthal\\
Leibniz Str.\ 10,
D-38678 Clausthal, Federal Republic of Germany

\end{center}

\vspace{1 cm}

\begin{abstract}
It is is explained why physical consistency requires substituting linear
observables by nonlinear ones for quantum systems with nonlinear time
evolution of pure states. The exact meaning and the concrete physical
interpretation are described in full detail for a special case of the
nonlinear Doebner-Goldin equation.
\end{abstract}

\vspace{1 cm}

\section{Schr\"odinger dynamics} \label{SDGND}

By \notion{Schr\"odinger dynamics} we mean a strongly continuous
two-parameter family of mappings $\hat\beta_{t_2,t_1}$ of some Hilbert
space $\hilb H$ of (pure) states onto itself, defined for
$t_1,t_2\in\bbbr$ and satisfying the following conditions:
\begin{eqnarray}
\hat\beta_{t,t}(\Psi)&=&\makebox[2cm][l]{$\Psi$}\quad
\forall\,t\in\bbbr\,,\,\Psi\in\hilb H\,, \label{beta}\\
\hat\beta_{t_3,t_2}\Bigl(\hat\beta_{t_2,t_1}(\Psi)\Bigr) &=&
\makebox[2cm][l]{$\hat\beta_{t_3,t_1}(\Psi)$}
\quad\forall\,t_1,t_2,t_3\in\bbbr
\,,\,\Psi\in\hilb H\setminus \left\{ 0 \right\} \,, \label{grp} \\
\hat\beta_{t_2,t_1}(c\Psi) &=&
\makebox[2cm][l]{$c\hat\beta_{t_2,t_1}(\Psi)$} \quad
\forall\,t_1,t_2\in\bbbr \,,\,c\in\bbbc\,,\,
\Psi\in\hilb H\setminus \left\{0\right\}\,.\label{homo}
\end{eqnarray}
In order to be consistent with the standard (nonrelativistic)
interpretation
\begin{equation} \label{guide1}
\begin{array}[c]{rcl}
\rho_{\Psi_t}&\stackrel{\rm def}{=}&\modulus{\Psi_t}^2 \\
&=& \mbox{probability density for particle position at time } t\,,
\end{array}
\end{equation}
we add the requirement
\begin{equation} \label{NC}
\norm{\hat\beta_{t_2,t_1}(\Psi)} = \norm{\Psi} \quad
\forall\,t_1,t_2\in\bbbr\,\,\Psi\in\hilb H\,.
\end{equation}
In other words:\footnote{In order to allow, e.g., for the
Bialynicki-Birula equation \cite{BiBiMy},
condition (\ref{homo}) could be generalized to
$$
\hat\beta_{t_2,t_1}(c\Psi)=c\,e^{i\varphi(c,t_1,t_2)}\,
\hat\beta_{t_2,t_1}(\Psi)\quad\forall\,t_1,t_2\in\bbbr
\,,\,c\in\bbbc\,,\, \Psi\in\hilb H\setminus \left\{0\right\}\,,
$$
where  $\varphi$ is some real-valued function.
Since we are mainly interested in the Doebner-Goldin equation, here,
this generalization is not necessary. The  physical importance of
(\ref{homo}) was  extensively discussed in \cite{KibbleNLS}.}
\begin{quote}
A Schr\"odinger dynamics is a norm conserving
propagator fulfilling (\ref{homo}).
\end{quote}
\vskip 4mm

\noindent
We call a Schr\"odinger dynamics $\left\{\hat\beta_{t_2,t_1}\right\}$
{\em linear\/} if
\begin{equation} \label{WIGNER}
\modulus{\left\langle
\frac{\hat\beta_{t_2,t_1}(\Phi)}{\norm{\hat\beta_{t_2,t_1}(\Phi)}} \bigg|
\frac{\hat\beta_{t_2,t_1}(\Psi)}{\norm{\hat\beta_{t_2,t_1}(\Psi)}}
\right\rangle} =
\modulus{\left\langle \frac{\Phi}{\norm{\Phi}} \bigg|
\frac{\Psi}{\norm{\Psi}}
\right\rangle}\quad\forall\,t_1,t_2\in\bbbr\,,\,\Psi,\Phi\in\PH
\end{equation}
holds. Otherwise it is called {\em nonlinear\/}.
Actually, since (\ref{beta}) and (\ref{grp}) imply
$$
\hat\beta_{t_1,t_2} = \hat\beta^{-1}_{t_2,t_1} \quad
\forall\,t_1,t_2\in\bbbr\,,
$$
a Schr\"odinger dynamics is given by the one-parameter family of
invertible norm conserving mappings
$\beta_t\stackrel{\rm def}{=}\hat\beta_{t,0}:$
$$
\hat\beta_{t_2,t_1} = \hat\beta_{t_2,0}\circ\hat\beta^{-1}_{t_1,0} \quad
\forall\,t_1,t_2\in\bbbr\,.
$$
Typically, such a
family is fixed by some nonlinear Schr\"odinger equation
\begin{equation} \label{nlseq}
i\hbar\frac{\partial}{\partial t}\beta_t(\Psi)=\Op
H_t\Bigl(\beta_t(\Psi)\Bigr) \,,
\end{equation}
with a suitable nonlinear Hamiltonian $\Op H_t$ fulfilling
\begin{equation} \label{homham}
\Op H_t(c\Psi) = c \Op H_t(\Psi)\quad\forall\,c\in\bbbc
\end{equation}
on a suitable dense set of state vectors. For example,
\begin{equation} \label{DGG}
H_t(\Psi) =\left(-\frac{\hbar^2}{2m}\Delta+V(\xv,t)\right)
\Psi + iD\hbar G(\Psi)\;, \quad G(\Psi) \stackrel{\rm def}{=}
\Nv^2\Psi + \modulus{\frac{\Nv\Psi}{\Psi}}^2 \Psi
\end{equation}
was considered in \cite{DoGo} and generalized\footnote{Equation
(\ref{DGG}) is the special case $c_1=-1\,,\; c_2=\ldots=c_5=0$ of
(\ref{genDGG}).} to
\begin{equation} \label{genDGG}
\begin{array}[c]{rcl}
H_t(\Psi) &=& \left(-\frac{\hbar^2}{2m}\Delta+V(\xv,t)\right)
\Psi + i\frac{\hbar D}{2} \frac{\Delta\rho_\Psi}{\rho_\Psi}\Psi\\
&& + \rule{0mm}{6mm}\hbar D\left( c_1 \frac{\Nv\cdot \vec
J_\Psi}{\rho_\Psi} + c_2 \frac{\Delta\rho_\Psi}{\rho_\Psi} + c_3
\frac{\vec J_\Psi^2}{\rho_\Psi^2}+ c_4 \frac{\vec J_\Psi\cdot
\Nv\rho_\Psi}{\rho_\Psi^2} + c_5 \frac{(\Nv\rho_\Psi)^2}{\rho_\Psi^2}
\right) \Psi
\end{array}
\end{equation}
where
$$
\rho_\Psi \stackrel{\rm def}{=} \modulus\psi^2\;,\quad \vec J_\Psi
\stackrel{\rm def}{=} \frac{1}{2i} \left( \overline{\Psi} \Nv \Psi -
\Psi \Nv \overline{\Psi} \right)
$$
\cite{DoGoGen}. For
\begin{equation} \label{NLTE}
c_1=1\;, \quad c_2+2c_5=0\;,\quad c_3=0\;,\quad c_4=-1\,,
\end{equation}
the linear dynamics $\beta_{0,t}\,$, characterized by\footnote{For
time-independent $V$ we have $\beta_{0,t}=e^{-\frac i\hbar\Op H_{0,0}t}\,$,
of course.}
\begin{equation} \label{lindyn}
i\hbar\partial_t \, \beta_{0,t}\Psi = \Op H_{0,t} \, \beta_{0,t}\Psi\;,
\quad  \Op H_{0,t}=-\frac{\hbar^2}{2m}\Delta+V(\xv,t)\,,
\end{equation}
is {\em affiliated\/} to $\beta_t$ by some nonlinear norm conserving
intertwining operator $\Op N:$
$$
\beta_t \circ \Op N = \Op N\circ\beta_{0,t}
$$
(see \cite{NattT93}). Let us restrict to the case
$c_2=-\frac{mD}{\hbar}\,$. Then this intertwiner is given by $\Op N=\Op
N_{\!D}\,$, where
\begin{equation} \label{nlint}
\left(\Op N_{\! D}(\Psi)\right)(\xv) \stackrel{\rm def}{=}
\cases{e^{i\frac{mD}{\hbar}\ln\rho_\Psi(\xv)}\Psi(\xv)&
if $\Psi(\xv)$ defined and $\ne 0$\cr 0&else}\,.
\end{equation}
(see \cite[Sect.\ 3.4]{NattD}). By Lebesque's bounded convergence
theorem \cite[p.\ 110]{Halmos} it is easily seen that (\ref{nlint})
defines a strongly continuous norm conserving mapping $\Op N_{\! D}$ from
$\hilb H$ onto itself with inverse
\begin{equation} \label{invint}
\left({\Op N_{\! D}}\right)^{-1} = \Op N_{\! -D} \,.
\end{equation}
Actually (\ref{nlint}), (\ref{nlseq}) and\footnote{Note, however, that
$$
\Op H_{D,t} \stackrel{\rm def}{=} \left(i\hbar\frac{{\rm d}}{{\rm
d}t}\beta^{-1}_{D,t}\right) \circ \beta_{D,t} \ne\Op N_{\! D} \circ \Op
H^0_t \circ \Op N_{\! -D}\,,
$$
as can be easily checked when $\Psi(\xv) = e^{-x}$ in some open region,
even though
$$
\left\langle \Psi \mid \Op H_{D,t}(\Psi) \right\rangle = \left\langle \Op
N_{\! -D}(\Psi) \mid \Op H^0_t \Op N_{\! -D}(\Psi) \right\rangle\,.
$$
What about the relation between $\norm{\Op H_{D,t}(\Psi)}$ and $\norm{\Op
H_0 \Op N_{\! -D}(\Psi)}$ ?}
\begin{equation} \label{itr}
\beta_t = \beta_{D,t} \stackrel{\rm def}{=} \Op N_{\! D}\circ
\beta_{0,t} \circ \Op N_{\! -D}
\end{equation}
should be taken as a definition for (\ref{genDGG}) when $c_1=-c_4=1$ and
$c_2=c_3=c_5=0\,$.
\medskip

\section{Consequences of nonlinearity} \label{SCNL}

Let us denote by $\SQL$ the set of all orthogonal projections in $\hilb
H:$
$$
\SQL \stackrel{\rm def}{=} \left\{ \Op P\in\hilb B(\hilb H): \Op P=\Op
P^*\Op P \right\}\,.
$$
Then the following is well known (see, e.g., \cite[\S 3--2]{Piron}).
\vskip 5mm

\begin{thm}[Wigner] \label{TWigner}
A Schr\"odinger dynamics $\left\{\beta_t\right\}_{t\in\bbbr}$ on $\hilb
H\,,$ where ${\rm dim\,}\hilb H \geq 3\,,$ is linear
if and only if for every $t\in\bbbr$ there is either a unitary or an
anti-unitary\footnotemark operator $\OU_t$ with
$$
\Op P_{\beta_t(\Psi)} = \Op P_{\OU_t\Psi} = \OU_t \Op P_\Psi
\OU_t^*\quad\forall\, \Psi\in\hilb H\,,
$$
where
$$
\Op P_\Psi\,\Phi \stackrel{\rm def}{=} \norm\Psi^{-2}\left\langle \Psi
\mid \Phi \right\rangle \Psi\quad \mbox{for } \Phi\in\hilb
H\,,\,\Psi\in\PH\,.
$$
\end{thm}
\footnotetext{\rm Actually, thanks to strong continuity of $\beta_t$ and
$\beta_0=\Op 1\,$, $\OU_t$ cannot be anti-unitary. But this is of no
relevance here.}
\vskip 5mm

\noindent
Perhaps less well-known is the following.

\begin{thm} \label{Tmix}
A Schr\"odinger dynamics $\left\{\beta_t\right\}_{t\in\bbbr}$ is linear
if and only the following statement is correct:

\noindent
Let
$$
\sum_{\nu=0}^\infty \underbrace{\lambda_\nu}_{\geq 0} = \sum_{\nu'=0}^\infty
\underbrace{\lambda'_{\nu'}}_{\geq 0} = 1 \;;\; \quad
\left\{\Psi_\nu\right\},\left\{\Psi'_{\nu'}\right\}\subset\PH\,.
$$
Then
\begin{equation} \label{mixok}
\sum_{\nu=0}^\infty \lambda_\nu\omega_{\Psi_\nu}(\Op P) =
\sum_{\nu'=0}^\infty \lambda'_{\nu'}\omega_{\Psi'_{\nu'}}(\Op P)
\Longrightarrow \sum_{\nu=0}^\infty
\lambda_\nu\omega_{\beta_t(\Psi_\nu)}(\Op P) = \sum_{\nu'=0}^\infty
\lambda'_{\nu'}\omega_{\beta_t(\Psi'_{\nu'})}(\Op P)
\end{equation}
holds for all $t\in\bbbr$ and $\Op P\in\SQL\,$,where
$$
\omega_\Psi(\Op P) \stackrel{\rm def}{=} \frac{\left\langle \Psi \mid
\Op P\Psi \right\rangle}{\left\langle \Psi \mid \Psi \right\rangle}
\mbox{ for } \Psi\in\PH \,,\, \Op P\in\SQL\,.
$$
\end{thm}
\vskip 5mm

{\bf Proof of Theorem \ref{Tmix}:}
As usual, denote by $\SS$ the set of trace class operators on $\hilb H\,$.
Assume (\ref{mixok}). Then
$$
\begin{array}[c]{c}
\alpha_t(\Op T) \stackrel{\rm def}{=} \sum_{\nu=0}^\infty \lambda_\nu \Op
P_{\beta_t(\Psi_\nu)} \\
\mbox{for } \Op T=\sum_{\nu=0}^\infty \lambda_\nu \Op
P_{\Psi_\nu} \in \SSP \stackrel{\rm def}{=}
\left\{ \Op T\in\SS: \, \Op T \geq 0\,, \; \trace(\Op T)=1\right\}
\end{array}
$$
is a consistent definition. For $\Op T\in\SS$
let us denote by $\pm\Op T_\pm$ its positive resp.\ negative part:
$$
\Op T=\Op T_+ - \Op T_-\;; \quad \Op T_+,\Op T_- \geq\Op 0\,.
$$
Then the consistent extension
$$
\alpha_t(\Op T) \stackrel{\rm def}{=} \trace(\Op T_+)\,
\alpha_t\left(\frac{\Op T_+}{\trace(\Op T_+)}\right) - \trace(\Op T_-)\,
\alpha_t\left(\frac{\Op T_-}{\trace(\Op T_-)}\right) \quad \mbox{for }
t\in\bbbr\,,\,\Op T\in\SS\,.
$$
defines a family of invertible linear
mappings $\alpha_t:\, \SS \longrightarrow\SS\,$, mapping $\SSP$ onto
itself. Such mappings are known to be implemented by either unitary or
anti-unitary operators $U_T:$
$$
\Op P_{\beta_t(\Psi)} = \alpha_t(\Op P_\Psi) = \OU_t \Op P_\Psi \OU_t^*
\quad \forall\,  \Psi\in\PH
$$
(see \cite[Corollary 2.3.2]{Davies}). By (the easy part of) Wigner's
theorem this shows that $\left\{\beta_t\right\}$ is linear. Conversely,
(\ref{mixok}) follows from linearity of the Schr\"odinger dynamics by
(the nontrivial part of) Wigner's theorem.\quad\rule[-2mm]{2mm}{4mm}
\vskip 5mm

\begin{lemma} \label{Lnleq}
A Schr\"odinger dynamics $\left\{\beta_t\right\}_{t\in\bbbr}$ is linear
if and only if one of the following (equivalent) statements is correct:
\begin{itemize}
\item[(i)]
The following \Notion{primitive causality}
condition holds:
\begin{quote}
Let $t\in\bbbr$ and $\Op P\in\SQL\,$. Then there is a $\Op P'\in\SQL$
fulfilling
$$
\omega_\Psi(\Op P) = \omega_{\beta_t(\Psi)}(\Op P') \quad
\forall\,\Psi\in \hilb H\,.
$$
\end{quote}
\item[(ii)]
There is a \Notion{Heisenberg picture} in the usual sense:
\begin{quote}
Let $t\in\bbbr$ and $\Op P\in\SQL\,$. Then there is a $\Op P(t)\in\SQL$
fulfilling
$$
\omega_{\beta_t(\Psi)}(\Op P) = \omega_\Psi\Bigl(\Op P(t)\Bigr)
\quad \forall\,\Psi\in \hilb H\,.
$$
\end{quote}
\item[(iii)]
Let $t\in\bbbr$ and $\Op P, \Op P'\in\SQL\,$. Then
$$
\omega_\Psi(\Op P) = 1 \;
\Longleftrightarrow \; \omega_{\beta_t(\Psi)}(\Op P') = 1 \quad
\forall\,\Psi\in \hilb H\,.
$$
implies
$$
\omega_\Psi(\Op P) = 0 \;
\Longleftrightarrow \; \omega_{\beta_t(\Psi)}(\Op P') = 0 \quad
\forall\,\Psi\in \hilb H\,.
$$
\end{itemize}
\end{lemma}
\vskip 5mm

\section{Faster than light signals?} \label{Sentangle}

If the Schr\"odinger dynamics $\left\{\beta_t\right\}_{t\in\bbbr}$ is
nonlinear, then (\ref{mixok}) does not hold for all $t\in\bbbr$ and $\Op
P\in\SQL\,$. This can be understood as a warning that causality
problems (faster than light signals) might arise in a nonlinear
theory.\footnote{See \cite{Gisin2} and also Gisin's contribution to
these proceedings.} Let us have a qualitative discussion of this problem
without any assumption concerning the measuring process.
\medskip

\noindent
To have a simple model let us begin with standard Schr\"odinger theory
of one particle in an external potential $V(\xv,t)\,$, i.e.\
time-evolution is given by some unitary propagator $\Op U_t$ depending on
$V(\xv,t)\,$:
$$
i\hbar\partial_t \OU_t \Phi =
\left(-\frac{\hbar^2}{2m}\Delta_\xv+V(\xv,t)\right) \OU_t \Phi\,.
$$
Now consider the nonlinear Schr\"odinger dynamics on $\hilb
H=L^2(\bbbr^3)\otimes L^2(\bbbr^3)$ defined by
(\ref{itr}), where
$$
\beta_{0,t} = \Op U_t \otimes \Op U_t\,,
$$
$\Op N_{\!D}$ being always given by (\ref{nlint}) whatever the dimension
of $\xv$-space may be:
$$
\Op N_{\!D}(\Psi) = e^{i\frac{mD}{\hbar}\ln\rho_\Psi}\Psi
\quad\mbox{almost everywhere}\,.
$$
Then for $\Phi_1,\Phi_2,\hat\Phi_1,\hat\Phi_2\in L^2(\bbbr^3)$
$$
\Psi_t \stackrel{\rm def}{=} \Op N_{\!D} \left(\sum_j \left(\Op U_t
\Phi_j\right) \otimes \left(\Op U_t \hat\Phi_j\right)\right)
$$
is a solution of the corresponding nonlinear Schr\"odinger equation
$$
i\hbar\partial_t \Psi_t =
\begin{array}[t]{l}
\left(-\frac{\hbar^2}{2m}\Delta_{\vec X}+V(\xv_1,t)
+V(\xv_2,t)\right)\Psi_t + i\frac{\hbar D}{2} \frac{\Delta_{\vec
X}\rho_{\Psi_t}}{\rho_{\Psi_t}}\Psi_t\\
+ \hbar D\left( \frac{\Nv_{\!\vec X}\cdot \vec J_{\Psi_t}}{\rho_{\Psi_t}}
-\frac{mD}{\hbar} \frac{\Delta\rho_\Psi}{\rho_\Psi} - \frac{\vec
J_{\Psi_t}\cdot \Nv_{\!\vec X}\rho_{\Psi_t}}{\rho_{\Psi_t}^2} +
\frac{mD}{2\hbar} \frac{(\Nv\rho_\Psi)^2}{\rho_\Psi^2} \right)
\Psi_t\;,\quad \vec X =(\xv_1,\xv_2)\,,
\end{array}
$$
for which we assume \ref{guide1} as in ordinary two-particle Schr\"odinger
theory. Even though $\Op N_{\! D}$ is nonlinear, we always have
\begin{equation} \label{pseudolin}
\Psi_j\Psi_k = 0 \mbox{ for } j\ne k \; \Longrightarrow\; \Op
N_{\! D}\left(\sum_k c_k \Psi_k\right) = c_k \sum_k \Op N_{\! D}(\Psi_k)\,.
\end{equation}
Moreover, we have the separability property
\begin{equation} \label{separab}
\Op N_{\! D}(\Phi\otimes\Psi) = \Op N_{\! D}(\Phi)\otimes\Op N_{\!
D}(\Psi) \,.
\end{equation}
Let us assume that the $\OU_t\Phi_j$ are (essentially) localized in the
`laboratory' and that their supports are (essentially) disjoint. Then, by
(\ref{pseudolin}) and (\ref{separab})
we have
$$
\Psi_t =  \sum_j \Op N_{\!D} \left(\Op U_t
\Phi_j\right) \otimes \Op N_{\!D} \left(\Op U_t \hat\Phi_j\right)
$$
and, consequently,
$$
\left\langle \Psi_t \mid (\OA\otimes\Op 1) \Psi_t \right\rangle =
\sum_{j,k} \lambda_{jk}(t) \left\langle \Op N_{\! D}(\OU_t\Phi_j) \mid
\OA \, \Op N_{\! D}(\OU_t\Phi_k)\right\rangle  \,,
$$
where
$$
\lambda_{jk}(t) \stackrel{\rm def}{=} \left\langle \Op N_{\!
D}(\OU_t\hat\Phi_j) \mid \Op N_{\! D}(\OU_t\hat\Phi_k) \right\rangle \,,
$$
is (essentially) valid.  Therefore, as long as the vectors $\Op N_{\!
D}(\OU_t\hat\Phi_k)$ are pairwise orthogonal the partial state with respect
to the `observables' $\OA\otimes\Op 1$ is (essentially) the mixed
state
\begin{equation} \label{mixe}
\omega_{\Psi_t}(\OA\otimes\Op 1) = \sum_j \lambda_{jj}\, \omega_{\Op N_{\!
D}(\OU_t\Phi_k)}(\OA) \,.
\end{equation}
Assume the supports of the $\OU_t\hat\Phi_k$ to be `behind the moon' and
initially pairwise disjoint.\footnote{Then (\ref{mixe}) is exact.}
So the $\Op N_{\! D}(\OU_t\hat\Phi_k)$ are initially pairwise
orthogonal.\footnote{Note that ${\rm supp\/}\Op N_{\! D}(\Psi) \subset
\Psi\,$.} Thanks to nonlinearity of $\Op N_{\! D}\,$, however, this
orthogonality can be (sufficiently) destroyed by applying a  suitable
exterior field `behind the moon' causing the (essential) supports of the
$\OU_t\hat\Phi_k$ to overlap:\footnote{For instance, if (for
$\xv=x\in\bbbr^1$) we define
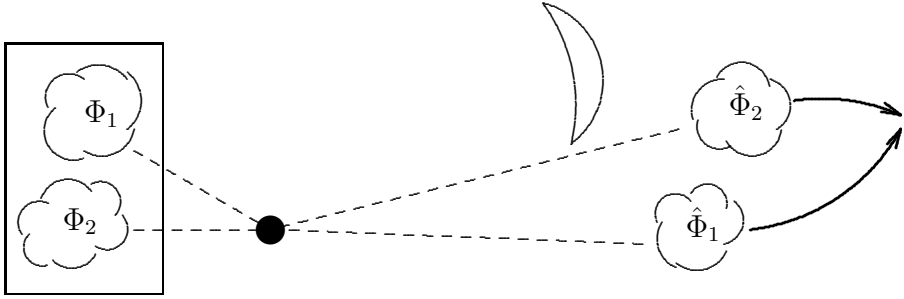
$$
\Phi_\pm(x) = \cases{\pm e^{\pm\frac{\hbar\pi}{4mD}}&for
$x\in(0,+1)\,$,\cr 1&for $x\in(-1,0)\,$,\cr 0&else\,,}
$$
we have
$$
\left\langle \Phi_- \mid \Phi_+ \right\rangle = 0\;, \quad
\frac{\left\langle \Op N_{\! D}(\Phi_-) \mid \Op N_{\! D}(\Phi_+)
\right\rangle}{\norm{\Phi_-} \, \norm{\Phi_+}} = \sqrt{\frac{2}{1 +
\cosh\left(\frac{\hbar\pi}{2mD}\right)}} \; \left( \approx 0.755 \mbox{
for } \frac{\hbar}{mD}=1\right)\,.
$$}
$$
\font\thinlinefont=cmr5
\begingroup\makeatletter\ifx\SetFigFont\undefined
% extract first six characters in \fmtname
\def\x#1#2#3#4#5#6#7\relax{\def\x{#1#2#3#4#5#6}}%
\expandafter\x\fmtname xxxxxx\relax \def\y{splain}%
\ifx\x\y   % LaTeX or SliTeX?
\gdef\SetFigFont#1#2#3{%
  \ifnum #1<17\tiny\else \ifnum #1<20\small\else
  \ifnum #1<24\normalsize\else \ifnum #1<29\large\else
  \ifnum #1<34\Large\else \ifnum #1<41\LARGE\else
     \huge\fi\fi\fi\fi\fi\fi
  \csname #3\endcsname}%
\else
\gdef\SetFigFont#1#2#3{\begingroup
  \count@#1\relax \ifnum 25<\count@\count@25\fi
  \def\x{\endgroup\@setsize\SetFigFont{#2pt}}%
  \expandafter\x
    \csname \romannumeral\the\count@ pt\expandafter\endcsname
    \csname @\romannumeral\the\count@ pt\endcsname
  \csname #3\endcsname}%
\fi
\fi\endgroup
\mbox{\beginpicture
\setcoordinatesystem units < 1.000cm, 1.000cm>
\unitlength= 1.000cm
\linethickness=1pt
\setplotsymbol ({\makebox(0,0)[l]{\tencirc\symbol{'160}}})
\setshadesymbol ({\thinlinefont .})
\setlinear
%
% Fig CIRCULAR ARC object
%
\linethickness= 0.500pt
\setplotsymbol ({\thinlinefont .})
\circulararc 106.260 degrees from  1.873 24.987 center at  1.587 25.368
%
% Fig CIRCULAR ARC object
%
\linethickness= 0.500pt
\setplotsymbol ({\thinlinefont .})
\circulararc 112.620 degrees from  1.302 24.701 center at  1.540 25.067
%
% Fig CIRCULAR ARC object
%
\linethickness= 0.500pt
\setplotsymbol ({\thinlinefont .})
\circulararc 163.740 degrees from  0.921 25.273 center at  1.159 24.967
%
% Fig CIRCULAR ARC object
%
\linethickness= 0.500pt
\setplotsymbol ({\thinlinefont .})
\circulararc 157.380 degrees from  1.302 25.654 center at  1.102 25.416
%
% Fig CIRCULAR ARC object
%
\linethickness= 0.500pt
\setplotsymbol ({\thinlinefont .})
\circulararc 90.000 degrees from  1.873 25.654 center at  1.492 25.368
%
% Fig TEXT object
%
\put{$\Phi_1$} [lB] at  1.302 25.082
%
% Fig CIRCULAR ARC object
%
\linethickness= 0.500pt
\setplotsymbol ({\thinlinefont .})
\circulararc 162.510 degrees from  1.715 23.844 center at  1.524 24.026
%
% Fig CIRCULAR ARC object
%
\linethickness= 0.500pt
\setplotsymbol ({\thinlinefont .})
\circulararc 166.004 degrees from  1.460 23.431 center at  1.610 23.659
%
% Fig CIRCULAR ARC object
%
\linethickness= 0.500pt
\setplotsymbol ({\thinlinefont .})
\circulararc 107.492 degrees from  0.889 23.241 center at  1.140 23.498
%
% Fig CIRCULAR ARC object
%
\linethickness= 0.500pt
\setplotsymbol ({\thinlinefont .})
\circulararc 151.928 degrees from  0.572 23.590 center at  0.750 23.416
%
% Fig CIRCULAR ARC object
%
\linethickness= 0.500pt
\setplotsymbol ({\thinlinefont .})
\circulararc 149.487 degrees from  0.826 24.003 center at  0.707 23.749
%
% Fig CIRCULAR ARC object
%
\linethickness= 0.500pt
\setplotsymbol ({\thinlinefont .})
\circulararc 126.208 degrees from  1.397 24.162 center at  1.088 23.945
%
% Fig TEXT object
%
\put{$\Phi_2$} [lB] at  1.016 23.654
%
% Fig CIRCULAR ARC object
%
\linethickness= 0.500pt
\setplotsymbol ({\thinlinefont .})
\circulararc 110.607 degrees from  9.811 25.654 center at  9.791 25.241
%
% Fig CIRCULAR ARC object
%
\linethickness= 0.500pt
\setplotsymbol ({\thinlinefont .})
\circulararc 136.018 degrees from  9.462 25.114 center at  9.754 25.029
%
% Fig CIRCULAR ARC object
%
\linethickness= 0.500pt
\setplotsymbol ({\thinlinefont .})
\circulararc 144.727 degrees from  9.842 24.829 center at 10.182 25.024
%
% Fig CIRCULAR ARC object
%
\linethickness= 0.500pt
\setplotsymbol ({\thinlinefont .})
\circulararc 137.000 degrees from 10.509 24.987 center at 10.330 25.302
%
% Fig CIRCULAR ARC object
%
\linethickness= 0.500pt
\setplotsymbol ({\thinlinefont .})
\circulararc 141.773 degrees from 10.414 25.559 center at 10.043 25.501
%
% Fig TEXT object
%
\put{$\hat\Phi_2$} [lB] at  9.874 25.178
%
% Fig CIRCULAR ARC object
%
\linethickness= 0.500pt
\setplotsymbol ({\thinlinefont .})
\circulararc 163.150 degrees from  9.842 23.940 center at  9.590 23.999
%
% Fig CIRCULAR ARC object
%
\linethickness= 0.500pt
\setplotsymbol ({\thinlinefont .})
\circulararc 131.545 degrees from  9.811 23.463 center at  9.746 23.778
%
% Fig CIRCULAR ARC object
%
\linethickness= 0.500pt
\setplotsymbol ({\thinlinefont .})
\circulararc 106.260 degrees from  9.176 23.209 center at  9.450 23.539
%
% Fig CIRCULAR ARC object
%
\linethickness= 0.500pt
\setplotsymbol ({\thinlinefont .})
\circulararc 100.387 degrees from  8.985 23.749 center at  9.332 23.527
%
% Fig CIRCULAR ARC object
%
\linethickness= 0.500pt
\setplotsymbol ({\thinlinefont .})
\circulararc 165.882 degrees from  9.366 24.035 center at  9.162 23.864
%
% Fig TEXT object
%
\put{$\hat\Phi_1$} [lB] at  9.303 23.559
%
% Fig CIRCULAR ARC object
%
\linethickness=1pt
\setplotsymbol ({\makebox(0,0)[l]{\tencirc\symbol{'160}}})
%
% arrow head
%
\plot 11.902 25.221 12.160 25.178 11.953 25.337 /
\circulararc 31.891 degrees from 12.160 25.178 center at 11.113 22.773
%
% Fig CIRCULAR ARC object
%
\linethickness=1pt
\setplotsymbol ({\makebox(0,0)[l]{\tencirc\symbol{'160}}})
%
% arrow head
%
\plot 12.093 24.734 12.160 24.987 11.982 24.796 /
\circulararc 54.658 degrees from 10.160 23.654 center at  9.870 26.256
%
% Fig CIRCULAR ARC object
%
\linethickness= 0.500pt
\setplotsymbol ({\thinlinefont .})
\circulararc 129.307 degrees from  7.779 24.797 center at  7.145 25.643
%
% Fig CIRCULAR ARC object
%
\linethickness= 0.500pt
\setplotsymbol ({\thinlinefont .})
\circulararc 49.550 degrees from  7.779 24.797 center at  5.559 25.321
%
% Fig ELLIPSE
%
\linethickness= 0.500pt
\setplotsymbol ({\thinlinefont .})
\put{\makebox(0,0)[l]{\circle*{ 0.381}}}
at  3.778 23.654
%
% Fig POLYLINE object
%
\linethickness= 0.500pt
\setplotsymbol ({\thinlinefont .})
\setdashes < 0.1270cm>
\plot  1.968 24.701  3.778 23.654 /
%
% Fig POLYLINE object
%
\linethickness= 0.500pt
\setplotsymbol ({\thinlinefont .})
\plot  1.968 23.654  3.778 23.654 /
%
% Fig POLYLINE object
%
\linethickness= 0.500pt
\setplotsymbol ({\thinlinefont .})
\plot  9.207 24.987  3.778 23.654 /
%
% Fig POLYLINE object
%
\linethickness= 0.500pt
\setplotsymbol ({\thinlinefont .})
\plot  3.778 23.654  8.731 23.463 /
%
% Fig POLYLINE object
%
\linethickness= 0.500pt
\setplotsymbol ({\thinlinefont .})
\setsolid
\putrectangle corners at  0.254 26.130 and  2.349 22.797
\linethickness=0pt
\putrectangle corners at  0.254 26.670 and 12.160 22.797
\endpicture}
$$
This way the partial state with respect to the operators $\OA\otimes\Op
1$ changes in a way not depending on the distance of the `moon' from the
`laboratory'.
\begin{quote}
Clearly things can be arranged\footnote{In principle these
considerations can be made numerically precise.} such as to produce {\bf
faster than light signals} in a realistic way, {\bf if all linear
observables can really be measured}.
\end{quote}

Admittedly, we used the additional assumption, that $\Psi_t$ describes a
state that can experimentally prepared.
\pagebreak

\section{Generalized projection valued measures} \label{SGPVM}

The essential message of Theorem \ref{Tmix}, Lemma \ref{Lnleq} and the
discussion in Section \ref{Sentangle} is twofold:
\begin{itemize}
\item[(i)]
For a nonlinear theory {\bf not all} $\Op P\in\SQL$ should be considered
as actually {\bf measurable} (in principle).
\item[(ii)]
For a nonlinear theory one should {\bf add} some kind of {\bf nonlinear
observables} to identify the initial conditions of classical mixtures and
to restore primitive causality as well as the Heisenberg picture.
\end{itemize}
\vskip 5mm

\begin{defin} \label{DGPV}
A one-parameter family $\OA$ of mappings $\Op E_B:\, \hilb H
\longrightarrow\hilb H\,$, defined for all Borel sets $B\subset\bbbr$ is
called a \Notion{generalized projection valued measure\,}\footnote{Maybe
this is not a good terminology, because it is only the probabilities
that are additive.} {\em (GPVM)}, if the following requirements are
fulfilled:
\begin{itemize}
\item[(i)]
For every $\Psi\in\PH$
$$
B \longmapsto \mu^{\Op A}_\psi(B) \stackrel{\rm def}{=} \omega_\Psi(\Op
E_B) = \frac{\norm{\Op E_B(\Psi)}^2}{\norm{\Psi}^2}
$$
defines a probability measure $\mu^\OA_\psi$ on $\bbbr$.
\item[(ii)]
For every pair of Borel sets $B_1,B_2\subset\bbbr:$
$$
\Op E_{B_1} \circ \Op E_{B_2} = \Op E_{B_1\cap B_2}\,.
$$
\item[(iii)]
For every $\Psi\in\PH$ and for every Borel set $B\in\bbbr:$
$$
\mu^{\Op A}_\psi(B) = 1 \; \Longrightarrow\; \Op E_B(\Psi)=\Psi\,.
$$
\end{itemize}
\end{defin}
\vskip 5mm

\begin{defin} \label{Dnlsr}
A {\em GPVM} $\OA=\left\{\Op E_B\right\}$ is called an
\Notion{observable} if for every Borel set $B\subset\bbbr$ there is (in
principle) an experimental test with the following properties:
\begin{itemize}
\item[(i)]
The probability to get a positive answer if the quantum system is in the
state $\omega_\Psi$ is $\mu^\OA_\Psi(B)\,$.
\item[(ii)]
The effect of such a test on the quantum system is described by the
instantaneous change
$$
\Psi \longmapsto \Op E_B(\Psi)\qquad \left(\mbox{wave packet collapse
\`a la L\"uders}\right)
$$
of the state vector $\Psi\,$.
\end{itemize}
An observable $\OA=\left\{\Op E_B\right\}$ is called \Notion{bounded}, if
$\Op E_C=\Op 1$ for some compact subset $C$ of $\bbbr\,$. It is called
\Notion{linear} if all the $\Op E_B$ are orthogonal projections.
The \Notion{expectation value} for an observable $\OA=\left\{\Op
E_B\right\}$ in the state characterized by $\Psi\in\hilb H$ is
$$
E_\Psi(\OA) \stackrel{\rm def}{=} \int \lambda\, {\rm
d}\mu^\OA_\psi(\lambda)\,.
$$
The observable is called \Notion{conserved} by the Schr\"odinger dynamics
$\beta_t$ if $\Op E_B\circ\beta_t = \beta_t\circ\Op E_B$ holds for all
$t\in\bbbr$ and for all Borel sets $B\subset\bbbr\,$.
\end{defin}
\vskip 5mm

\noindent
Let the quantum system be in the state $\omega_\Psi$ at time $t=0\,$.
Then the probability for getting a positive outcome for both a test at
time $t_1>0$ corresponding to $\Op E^{\OA_1}_{B_1}$ and a test at
time $t_2>t_1$ corresponding to $\Op E^{\OA_2}_{B_2}$ is
$$
\begin{array}[c]{rcl}
\mu^{\OA_1}_{\beta_{t_1}(\Psi)}(B_1) \,
\mu^{\OA_2}_{(\hat\beta_{t_2,t_1}\circ\Op
E^{\OA_1}_{B_1}\circ\beta_{t_1}) (\Psi)}(B_2) &=& \frac{\norm{\Op
E^{\OA_1}_{B_1}\left(\beta_{t_1}(\Psi)\right)}^2}
{\norm{\beta_{t_1}(\Psi)}^2}\, \frac{\norm{\left(\Op
E^{\OA_2}_{B_2}\circ\hat\beta_{t_2,t_1}\circ\Op
E^{\OA_1}_{B_1}\circ\beta_{t_1}\right)(\Psi)}^2}
{\norm{\left(\hat\beta_{t_2,t_1}\circ\Op E^{\OA_1}_{B_1}\circ\beta_{t_1}
\right)(\Psi)}^2}\\
&\stackscript{=}_{(\ref{NC})}& \frac{\norm{\left(\Op
E^{\OA_2}_{B_2}\circ \hat\beta_{t_2,t_1}\circ\Op
E^{\OA_1}_{B_1}\circ\beta_{t_1}\right)(\Psi)}^2} {\norm{\Psi}^2}\\
&\stackscript{\longrightarrow}_{t_1,t_2\to 0}& \frac{\norm{\left(\Op
E^{\OA_2}_{B_2}\circ\Op E^{\OA_1}_{B_1}\right)(\Psi)}^2} {\norm{\Psi}^2}\,.
\end{array}
$$

\noindent
Unfortunately, an observable $\Op E = \left\{ \Op E_B \right\}$ is not
uniquely characterized by its expectation values, unless it is
restricted to be linear. However:
\begin{quote}
``It can be maintained that all measurements are reducible to position
measurements (pointer readings).'' \cite{Nelson}
\end{quote}
This suggests that, in principle, the physical interpretation is already
completely fixed by the {\bf basic assumption} (\ref{guide1}), i.e.\ by the
identification
$$
\Op E^{\Op x^j}_B \cor \mbox{multiplication by } \chi_B(x^j)
$$
for the observable of the j-component of the position vector.
\medskip

\section{Physical identification of nonlinear observables} \label{Spinlo}

A typical measurement procedure for one-particle systems is as follows:
\begin{quote}
Apply exterior fields $F$ such that those particles for which $\OA$ has
a value in $B$ asymptotically ($t\to+\infty$) enter the time-dependent
region $\region O_t\subset \bbbr^3$ while the others are leaving it. Then
$$
\mu^\OA_B(\Psi) = \norm\Psi^{-2} \lim_{t\to+\infty} \norm{\chi_{{\cal
O}_t}\,\beta^F_t(\Psi)}^{2}\,,
$$
where $\beta^F_t$ denotes the Schr\"odinger dynamics adapted to the
applied fields.
\end{quote}
In this situation $\Op E^\OA_B(\Psi)$ should be identified with the initial
wave function that evolves like $\chi_{{\cal O}_t}\beta^F_t(\Psi)$ for
large $t:$
$$
\lim_{t\to+\infty} \norm{\beta^F_t\left(\Op
E^\OA_B(\Psi)\right) - \chi_{{\cal O}_t}\beta^F_t(\Psi)} =0\,,
$$
i.e.\
\begin{equation} \label{asbnlo}
E^\OA_B(\Psi) = \lim_{t\to+\infty} \left(\left(\beta^F_t\right)^{-1}
\circ \chi_{{\cal O}_t} \circ\beta^F_t\right)(\Psi)\,.
\end{equation}
\vskip 5mm

\noindent
As an example let us derive the nonlinear observable of linear momentum
for the theory defined by (\ref{itr})/(\ref{nlint}), now for
$\xv\in\bbbr^3$ and  with the additional restriction $V=0\,$.
Assuming momentum conservation (\ref{guide1}) implies\footnote{Actually,
$$
\lim_{t\to \infty} \int_{\frac tm
B} \modulus{\beta_{0,t}(\Psi)(\xv)}^2\,{\rm d}\xv = \int_B
\modulus{\tilde\Psi(\pv)}^2\, {\rm d}\pv
$$
is well-known to hold for $\beta_{0,t}=\exp\left(-\frac i\hbar\Delta
t\right)$ \cite[Sect.\ 15a]{Kemble}.}
\begin{equation} \label{defmp}
\mbox{probability for } \pv\in \tilde{\cal O} = \lim_{t\to \infty}
\int_{\frac tm \tilde{\cal O}} \modulus{\beta_{D,t}(\Psi)(\xv)}^2\,{\rm
d}\xv\,.
\end{equation}
Therefore, if $\Op p^1_D$ is the observable of linear momentum
in 1-direction, $\Op E^{\Op p^1_D}_B(\Psi)$ should coincide with that
initial wave function that evolves like $\chi_{\frac tm
B}(\Op x^1)\beta_{D,t}(\Psi)$ for large $t:$
$$
\lim_{t\to+\infty} \norm{\beta_{D,t}\left(\Op E^{\Op
p^1_D}_B(\Psi)(\Psi)\right) - \chi_{\frac tm B}(\Op
x^1)\beta_{D,t}(\Psi)} =0\,,
$$
i.e.\
\begin{equation} \label{nlmp}
\Op E^{\Op p^1_D}_B(\Psi) = \lim_{t\to+\infty}
\left(\beta_{D,-t}\circ \chi_{\frac tm B} \circ\beta_{D,t}\right)(\Psi)\,.
\end{equation}
 From (\ref{defmp}) we easily derive the following.\footnote{Note that
$$
\lim_{t\to+\infty}\norm{\left(\chi_{\frac tm B}\circ\beta_{D,t}\circ
\chi_{\frac tm  (\bbbr^3\setminus B)}(\frac \hbar i\Nv)\right)(\Psi)} = 0
$$
and
$$
\lim_{t\to+\infty}\norm{\left(\chi_{\frac tm B}\circ \beta_{D,t} \circ
\chi_{\frac tm B}(\frac \hbar i\Nv)\right)(\Psi) - \left(\beta_{D,t}
\circ \chi_{\frac tm B}(\frac \hbar i\Nv)\right)(\Psi)} = 0\,.
$$}
\vskip 5mm

\begin{lemma} \label{Lmompro}
For $\beta_{0,t}=\exp\left(-\frac i\hbar\Delta t\right)$ we have
$$
\lim_{t\to+\infty} \beta_{0,-t}\circ\chi_{\frac tm \tilde{\cal
O}}\circ\beta_{0,t} = \chi_{\tilde{\cal O}}(\frac \hbar i\Nv)
$$
in the strong operator topology.
\end{lemma}
\vskip 5mm

\noindent
Since, thanks to
$$
\Op N_{\!D}\circ\chi_{\frac tm\tilde{\cal O}} = \chi_{\frac tm\tilde{\cal
O}}\circ\Op N_{\!D}\,,
$$
we have
$$
\begin{array}[c]{rcl}
\beta_{D,-t}\circ\chi_{\frac tm\tilde{\cal O}}\circ\beta_{D,t} &=& \Op
N_{\!D}\circ\beta_{0,-t}\circ\Op N_{\!-D}\circ\chi_{\frac tm\tilde{\cal
O}}\circ\Op
N_{\!D}\circ\beta_{0,t}\circ\Op N_{\!-D}\\
&=& \Op N_{\!D}\circ\beta_{0,-t}\circ\chi_{\frac
tm\tilde{\cal O}}\circ\beta_{0,t}\circ\Op N_{\!-D}\,,
\end{array}
$$
Lemma \ref{Lmompro} shows that
$$
\Op E^{\Op p^1_D}_B(\Psi) = \mbox{s-}\lim_{t\to+\infty}
\beta_{D,-t}\circ\chi_{\frac tm B}\circ\beta_{D,t} = \Op
N_{\!D}\circ\chi_{\frac tmB}(\frac \hbar i\partial_1)\circ\Op N_{\!-D}\,,
$$
i.e.\ the (nonlinear) observable for linear momentum is
$$
\Op p^j_{D}= \left\{\Op N_{\!D}\circ\chi_B(\frac \hbar
i\partial_j)\circ\Op N_{\!-D}\right\} \quad \mbox{for } j=1,2,3\,.
$$
Even though
$$
E_{\beta_{D,t}(\Psi)}\left(\Op{\pv}_D\right) =
E_{\beta_{D,t}(\Psi)}\left(\Op{\pv}_0\right) = \frac{{\rm d}}{{\rm d}t}
\int \xv\,\rho_{\beta_{D,t}(\Psi)}(\xv)\,{\rm d}\xv \,,
$$
$\Op{\pv}_D$ and $\Op{\pv}_0$ -- contrary to what Weinberg
guessed\footnote{The Doebner-Goldin (\ref{genDGG}) equation can be shown
to fit into Weinberg's framework \cite{WeinTQ} if $c_1=1\;, \quad
c_2+2c_5=0\;,\quad c_3=0\;,\quad c_4=-1\,$.} \cite[Sec.~2]{WeinTQ} -- do not
coincide for $D\ne0\,,$
\begin{quote}
$\Op{\pv}_D$ is conserved by $\left\{\beta_{D,t}\right\}$ but
$\Op{\pv}_0$ is not.
\end{quote}
Finally, note that
$$
B_1\cap B_2=\emptyset \; \Longrightarrow \;
\norm{\Op E^\OA_{B_1\cup B_2}(\Psi)}^2 = \norm{\Op
E^\OA_{B_1}(\Psi)}^2 + \norm{\Op E^\OA_{B_2}(\Psi)}^2\,,
$$
holds for every observable $\OA\,$, but:
$$
\begin{array}[c]{rcl}
B_1\cap B_2=\emptyset &\stackrel{\rm i.g.}{\not
\Longrightarrow} & \left\langle \Op E^{\Op p^1_D}_{B_1}(\Psi)
\mid \Op E^{\Op p^1_D}_{B_2}(\Psi) \right\rangle =0\,,\\
B_1\cap B_2=\emptyset &\stackrel{\rm i.g.}{\not
\Longrightarrow} & \Op E^{\Op p^1_D}_{B_1\cup B_2}(\Psi)
= \Op E^{\Op p^1_D}_{B_1}(\Psi) + \Op E^{\Op p^1_D}_{B_2}(\Psi) \,.
\end{array}
$$
This fact is a reasonable consequence of nonlinearity of the
dynamics.
\medskip

\section{Consistent physical interpretation} \label{Scpi}

Let $\algebra A$ be the set of observables for the Schr\"odinger
dynamics $\left\{\beta_t\right\}$ and define
$$
\nlSQL \stackrel{\rm def}{=} \left\{ \Op
E^\OA_B:\, \OA\in\algebra A\,,\, \bbbr\supset B \mbox{ Borel}\right\}\,.
$$
Then we expect the following consistency conditions to be fulfilled:
\begin{itemize}
\item[(C1)]
Let $\Op E,\Op E'\in\nlSQL\,$. Then
$$
\omega_\Psi(\Op E) = \omega_\Psi(\Op E')\;\forall\,\Psi\in\PH\;
\Longrightarrow \; \Op E=\Op E'\,.
$$
\item[(C2)]
Enhanced by the semi-ordering $\prec\,$,
$$
\Op E_1\prec\Op E_2\;\stackrel{\rm def}{\Longleftrightarrow}\;
\omega_\Psi(\Op E_1) \leq \omega_\Psi(\Op E_2)\; \forall\,\Psi\in\PH
$$
and the ortho-complementation $\neg\,$, thanks to (C1) uniquely
characterized by
$$
\omega_\Psi\left(\neg\Op E\right) = 1 - \omega_\Psi(\Op
E)\quad \mbox{for } \Op E\in\nlSQL\,,\,\Psi\in \hilb H\,,
$$
$\nlSQL$ becomes a $\sigma$-complete orthomodular lattice ({\bf quantum
logic}).
\item[(C3)]
The {\bf pure states}\footnote{A \Notion{state} on
$\left(\nlSQL,\prec,\neg\right)$ is a $\sigma$-additive mapping $\omega:
\, \nlSQL\longrightarrow[0,1]$ with $\omega(\Op 1)=1\,$, fulfilling the
Jauch-Piron property
$$
\forall\,\Op E,\Op E'\in\nlSQL: \quad\omega(\Op E)=\omega(\Op E')=1
\Longrightarrow \omega(\Op E\land\Op E')=1\,.
$$
This guarantees for arbitrary $\Op E,\Op E'\in\nlSQL$ that
$$
\omega_\Psi(\Op E) =1 \Longleftrightarrow \omega_\Psi(\Op
E')=1\quad \forall\,\Psi\in\PH
$$
implies
$$
\omega_\Psi(\Op E) =0 \Longleftrightarrow \omega_\Psi(\Op
E')=0\quad \forall\,\Psi\in\PH \;.
$$} on
$\left(\nlSQL,\prec,\neg\right)$ are exactly
those of the form $\omega_\Psi\,,\;\Psi\in\PH\,$.
\item[(C4)]
The Schr\"odinger dynamics $\beta_{D,t}$ corresponds to a family of
automorphisms $\alpha_t$ of $\left(\nlSQL,\prec,\neg\right)\,$,
$$
\alpha_t(\Op E) = \beta_{D,-t}\circ\Op E\circ\beta_{D,t}\;\mbox{ for }
\Op E\in\nlSQL\,.
$$
\end{itemize}
\pagebreak

\begin{lemma} \label{Lallok}
Let $\algebra A$ be the set of observables for the Schr\"odinger
dynamics $\left\{\beta_t\right\}$ fulfilling conditions (C1)--(C4). Then:
\begin{itemize}
\item[(i)]
There are {\bf no problems with mixed states} :
$$
\begin{array}[c]{l}
\sum_{\nu=0}^\infty \lambda_\nu\omega_{\Psi_\nu}(\Op E) =
\sum_{\nu'=0}^\infty \lambda'_{\nu'}\omega_{\Psi'_{\nu'}}(\Op E)
\quad\forall\, \Op E\in\nlSQL\\
\Longrightarrow \sum_{\nu=0}^\infty
\lambda_\nu\omega_{\beta_t(\Psi_\nu)}(\Op E) = \sum_{\nu'=0}^\infty
\lambda'_{\nu'}\omega_{\beta_t(\Psi'_{\nu'})}(\Op
E)\quad\forall\,t\in\bbbr\,,\,\Op E\in\nlSQL\,.
\end{array}
$$
\item[(ii)]
{\bf Primitive causality} holds in the following sense:
\begin{quote}
For every $t\in\bbbr$ and for every $\Op E\in\nlSQL$ there is a $\Op
E_t\in\nlSQL$ fulfilling
$$
\omega_\Psi(\Op E) = \omega_{\beta_t(\Psi)}(\Op E_t)
\quad\forall\,\Psi\in \PH\,.
$$
\end{quote}
\item[(iii)]
There is a nonlinear \Notion{Heisenberg picture}:
\begin{quote}
Let $t\in\bbbr$ and $\Op E\in\nlSQL\,$. Then\footnote{Obviously $\Op
E(t)\cor\alpha_t(\Op E):$
$$
\underbrace{\omega_{\beta_{D,t}(\Psi)}(\Op
E)}_{\mbox{\scriptsize Schr\"odinger
picture}} = \underbrace{\omega_\Psi\left(\alpha_t(\Op
E)\right)}_{\mbox{\scriptsize Heisenberg picture}} \quad
\forall\,t\in\bbbr\,,\,\Psi\in\PH\,,\,\Op E\in\nlSQL\,.
$$} there is a $\Op E(t)\in\nlSQL$
fulfilling
$$
\omega_{\beta_t(\Psi)}(\Op E) =
\omega_\Psi\Bigl(\Op E(t)\Bigr) \quad \forall\,\Psi\in
\hilb H\,.
$$
\end{quote}
\end{itemize}
\end{lemma}
\vskip 5mm

\noindent
For the nonlinear Schr\"odinger dynamics given by
(\ref{itr})/(\ref{nlint}) a set of GPVM's fulfilling (C1)--(C4) is
\begin{equation} \label{specnlobs}
\algebra A = \left\{ \OA_{\!D}=\left\{\Op N_{\!D}\circ\chi_B(\OA)\circ\Op
N_{\!-D}\right\}: \, \OA \mbox{ self-adjoint}\right\}\,.
\end{equation}
Here $\left(\nlSQL,\prec,\neg\right)$ is isomorphic to the standard
quantum logic, the isomorphism $\gamma: \, \SQL\longrightarrow\nlSQL$
being implemented by $\Op N_{\!D}:$
$$
\gamma(\Op P) = \Op N_{\!D} \circ\Op P\circ \Op N_{\!-D} \; \mbox{ for }
\Op P\in\SQL\,.
$$
Since, by the correspondence
$$
\renewcommand{\arraystretch}{1.5}
\begin{tabular}{|l||l|l|}
\hline
&nonlinear theory&linear theory\\
\hline\hline
state vector&$\Op N_{\!D}(\Psi)$&$\Psi$\\
\hline
time-evolution&$\beta_{D,t}=\Op N_{\!D}
\circ\beta_{0,t}\circ\Op N_{\!-D}$&$\beta_{0,t}=e^{-\frac i\hbar\Op H^0_t
t}$\\
\hline
selected observables&$\OA_{\!D}=\left\{\Op
N_{\!D}\circ\chi_B(\OA)\circ\Op N_{\!-D}\right\}$&$\OA$\\
\hline
position observable&$\vec{\Op x}_{\!D}=\left\{\chi_B(\vec{\Op
x})\right\}$& $\vec{\Op x}=\,$multipl.\ by $\xv$\\
\hline
expectation values&$E_{\Op N_{\!D}(\Psi)}(\OA_{\!D})$&$E_\Psi(\OA)$\\
\hline
\end{tabular}
$$
our special nonlinear theory becomes physically equivalent
to standard linear quantum mechanics, we see:
\begin{quote}
A nonlinear theory need not a priori be less consistent than a linear one.
\end{quote}
\vskip 5mm

\noindent
{\sc\bf Challenge:}
\begin{quote}
Find a suitable Schr\"odinger dynamics $\left\{\beta_t\right\}$ (defining the
formal
singularities) for the original Doebner-Goldin equation (\ref{DGG}) and
a set $\algebra A$ of GPVM's respecting (C1)--(C4).
\end{quote}
\vskip 5mm

\section*{Acknowledgment}
I am indebted to H.-D. Doebner, G.A. Goldin and P. Nattermann for many
fruitful discussions.

\end{document}